\def\ref#1{\lbrack #1\rbrack}

\def\part#1#2{{\partial #1\over\partial #2}}

\def\arcsecf {\hbox{$.\!\!^{\prime\prime}$}}

\def\Real{{\rm I\mathchoice{\kern-0.70mm}{\kern-0.70mm}{\kern-0.65mm}%
  {\kern-0.50mm}R}}
\def\C{\rm C\kern-.42em\vrule width.03em height.58em depth-.02em
       \kern.4em}

\def\bx#1{\leavevmode\thinspace\hbox{\vrule\vtop{\vbox{\hrule\kern1pt
        \hbox{\vphantom{\tt/}\thinspace{\bf#1}\thinspace}}
      \kern1pt\hrule}\vrule}\thinspace}

{\catcode`\@=11
\gdef\SchlangeUnter#1#2{\lower2pt\vbox{\baselineskip 0pt \lineskip0pt
  \ialign{$\m@th#1\hfil##\hfil$\crcr#2\crcr\sim\crcr}}}
}

\def\lesssim{\mathrel{\mathpalette\SchlangeUnter<}}

\def\ueber#1#2{{\setbox0=\hbox{$#1$}%
  \setbox1=\hbox to\wd0{\hss$\scriptscriptstyle #2$\hss}%
  \offinterlineskip
  \vbox{\box1\kern0.4mm\box0}}{}}

\def\bx#1{\leavevmode\thinspace\hbox{\vrule\vtop{\vbox{\hrule\kern1pt
        \hbox{\vphantom{\tt/}\thinspace{\bf#1}\thinspace}}
      \kern1pt\hrule}\vrule}\thinspace}

\def\SFB{{This work was supported by the ``Sonderforschungsbereich
375-95 f\"ur
Astro--Teil\-chen\-phy\-sik" der Deutschen For\-schungs\-ge\-mein\-schaft.
RN was supported in part by NSF grant AST-9423209}}
 
\magnification=\magstep1
\input epsf
\voffset= 0.0 true cm
\vsize=19.8 cm     
\hsize=13.5 cm
\hfuzz=2pt
\tolerance=500
\abovedisplayskip=3 mm plus6pt minus 4pt
\belowdisplayskip=3 mm plus6pt minus 4pt
\abovedisplayshortskip=0mm plus6pt
\belowdisplayshortskip=2 mm plus4pt minus 4pt
\predisplaypenalty=0
\footline={\tenrm\ifodd\pageno\hfil\folio\else\folio\hfil\fi}

\def\la{\mathrel{\hbox{\rlap{\hbox{\lower4pt\hbox{$\sim$}}}\hbox{$<$}}}}
\def\ga{\mathrel{\hbox{\rlap{\hbox{\lower4pt\hbox{$\sim$}}}\hbox{$>$}}}}

\def\utw{\smash{\rlap{\lower5pt\hbox{$\sim$}}}}
\def\udtw{\smash{\rlap{\lower6pt\hbox{$\approx$}}}}

\def\getsto{\mathrel{\hbox{\rlap{$\gets$}\hbox{\raise2pt\hbox{$\to$}}}}}
\def\lid{\mathrel{\hbox{\rlap{\hbox{\lower4pt\hbox{$=$}}}\hbox{$<$}}}}
\def\gid{\mathrel{\hbox{\rlap{\hbox{\lower4pt\hbox{$=$}}}\hbox{$>$}}}}
\def\sol{\mathrel{\hbox{\rlap{\hbox{\raise4pt\hbox{$\sim$}}}\hbox{$<$}}}
}
\def\sog{\mathrel{\hbox{\rlap{\hbox{\raise4pt\hbox{$\sim$}}}\hbox{$>$}}}
}
\def\lse{\mathrel{\hbox{\rlap{\hbox{\raise4pt\hbox{$<$}}}\hbox{$\simeq$}
}}}
\def\gse{\mathrel{\hbox{\rlap{\hbox{\raise4pt\hbox{$>$}}}\hbox{$\simeq$}
}}}
\def\grole{\mathrel{\hbox{\lower2pt\hbox{$<$}}\kern-8pt
\hbox{\raise2pt\hbox{$>$}}}}
\def\leogr{\mathrel{\hbox{\lower2pt\hbox{$>$}}\kern-8pt
\hbox{\raise2pt\hbox{$<$}}}}
\def\loa{\mathrel{\hbox{\rlap{\hbox{\lower4pt\hbox{$\approx$}}}\hbox{$<$
}}}}
\def\goa{\mathrel{\hbox{\rlap{\hbox{\lower4pt\hbox{$\approx$}}}\hbox{$>$
}}}}

%
%

\font\kleinhalbcurs=cmmib10 scaled 833
\font\eightrm=cmr8
\font\sixrm=cmr6
\font\eighti=cmmi8
\font\sixi=cmmi6
\skewchar\eighti='177 \skewchar\sixi='177
\font\eightsy=cmsy8
\font\sixsy=cmsy6
\skewchar\eightsy='60 \skewchar\sixsy='60
\font\eightbf=cmbx8
\font\sixbf=cmbx6
\font\eighttt=cmtt8
\hyphenchar\eighttt=-1
\font\eightsl=cmsl8
\font\eightit=cmti8

\font\bxf=cmbx10
  \mathchardef\Gamma="0100
  \mathchardef\Delta="0101
  \mathchardef\Theta="0102
  \mathchardef\Lambda="0103
  \mathchardef\Xi="0104
  \mathchardef\Pi="0105
  \mathchardef\Sigma="0106
  \mathchardef\Upsilon="0107
  \mathchardef\Phi="0108
  \mathchardef\Psi="0109
  \mathchardef\Omega="010A
\def\rahmen#1{\vskip#1truecm}
\def\begfig#1cm#2\endfig{\par
\setbox1=\vbox{\rahmen{#1}#2}%
\dimen0=\ht1\advance\dimen0by\dp1\advance\dimen0by5\baselineskip
\advance\dimen0by0.4true cm
\ifdim\dimen0>\vsize\pageinsert\box1\vfill\endinsert
\else
\dimen0=\pagetotal\ifdim\dimen0<\pagegoal
\advance\dimen0by\ht1\advance\dimen0by\dp1\advance\dimen0by1.4true cm
\ifdim\dimen0>\vsize
\topinsert\box1\endinsert
\else\vskip1true cm\box1\vskip4true mm\fi
\else\vskip1true cm\box1\vskip4true mm\fi\fi}
\def\figure#1#2{\smallskip\setbox0=\vbox{\noindent\petit{\bf Fig.\ts#1.\
}\ignorespaces #2\smallskip
\count255=0\global\advance\count255by\prevgraf}%
\ifnum\count255>1\box0\else
\centerline{\petit{\bf Fig.\ts#1.\ }\ignorespaces#2}\smallskip\fi}

\def\xfigure#1#2#3#4{\midinsert\noindent
    $$\epsfxsize=#4truecm\epsffile{#3}$$
    \figure{#1}{#2}\endinsert}


\def\begtab#1cm#2\endtab{\par
\ifvoid\topins\midinsert\vbox{#2\rahmen{#1}}\endinsert
\else\topinsert\vbox{#2\kern#1true cm}\endinsert\fi}
\def\rahmen#1{\vskip#1truecm}
\def\begpet{\vskip6pt\bgroup\petit}
\def\endpet{\vskip6pt\egroup}
\def\begref{\par\bgroup\petit
\let\it=\rm\let\bf=\rm\let\sl=\rm\let\INS=N}
\def\petit{\def\rm{\fam0\eightrm}%
\textfont0=\eightrm \scriptfont0=\sixrm \scriptscriptfont0=\fiverm
 \textfont1=\eighti \scriptfont1=\sixi \scriptscriptfont1=\fivei
 \textfont2=\eightsy \scriptfont2=\sixsy \scriptscriptfont2=\fivesy
 \def\it{\fam\itfam\eightit}%
 \textfont\itfam=\eightit
 \def\sl{\fam\slfam\eightsl}%
 \textfont\slfam=\eightsl
 \def\bf{\fam\bffam\eightbf}%
 \textfont\bffam=\eightbf \scriptfont\bffam=\sixbf
 \scriptscriptfont\bffam=\fivebf
 \def\tt{\fam\ttfam\eighttt}%
 \textfont\ttfam=\eighttt
 \normalbaselineskip=9pt
 \setbox\strutbox=\hbox{\vrule height7pt depth2pt width0pt}%
 \normalbaselines\rm
\def\vec##1{\setbox0=\hbox{$##1$}\hbox{\hbox
to0pt{\copy0\hss}\kern0.45pt\box0}}}%
\let\ts=\thinspace
%
\font \tafontt=     cmbx10 scaled\magstep2
\font \tafonts=     cmbx7  scaled\magstep2
\font \tafontss=     cmbx5  scaled\magstep2
\font \tamt= cmmib10 scaled\magstep2
\font \tams= cmmib10 scaled\magstep1
\font \tamss= cmmib10
\font \tast= cmsy10 scaled\magstep2
\font \tass= cmsy7  scaled\magstep2
\font \tasss= cmsy5  scaled\magstep2
\font \tasyt= cmex10 scaled\magstep2
\font \tasys= cmex10 scaled\magstep1
\font \tbfontt=     cmbx10 scaled\magstep1
\font \tbfonts=     cmbx7  scaled\magstep1
\font \tbfontss=     cmbx5  scaled\magstep1
\font \tbst= cmsy10 scaled\magstep1
\font \tbss= cmsy7  scaled\magstep1
\font \tbsss= cmsy5  scaled\magstep1

\newbox\chsta\newbox\chstb\newbox\chstc
\def\centerpar#1{{\advance\hsize by-2\parindent
\rightskip=0pt plus 4em
\leftskip=0pt plus 4em
\parindent=0pt\setbox\chsta=\vbox{#1}%
\global\setbox\chstb=\vbox{\unvbox\chsta
\setbox\chstc=\lastbox
\line{\hfill\unhbox\chstc\unskip\unskip\unpenalty\hfill}}}%
\leftline{\kern\parindent\box\chstb}}
 \def \chap#1{
    \vskip24pt plus 6pt minus 4pt
    \bgroup
 \textfont0=\tafontt \scriptfont0=\tafonts \scriptscriptfont0=\tafontss
 \textfont1=\tamt \scriptfont1=\tams \scriptscriptfont1=\tamss
 \textfont2=\tast \scriptfont2=\tass \scriptscriptfont2=\tasss
 \textfont3=\tasyt \scriptfont3=\tasys \scriptscriptfont3=\tenex
     \baselineskip=18pt
     \lineskip=18pt
     \raggedright
     \pretolerance=10000
     \noindent
     \tafontt
     \ignorespaces#1\vskip7true mm plus6pt minus 4pt
     \egroup\noindent\ignorespaces}%
 \def \sec#1{
     \vskip25true pt plus4pt minus4pt
     \bgroup
 \textfont0=\tbfontt \scriptfont0=\tbfonts \scriptscriptfont0=\tbfontss
 \textfont1=\tams \scriptfont1=\tamss \scriptscriptfont1=\kleinhalbcurs
 \textfont2=\tbst \scriptfont2=\tbss \scriptscriptfont2=\tbsss
 \textfont3=\tasys \scriptfont3=\tenex \scriptscriptfont3=\tenex
     \baselineskip=16pt
     \lineskip=16pt
     \raggedright
     \pretolerance=10000
     \noindent
     \tbfontt
     \ignorespaces #1
     \vskip12true pt plus4pt minus4pt\egroup\noindent\ignorespaces}%
 \def \subs#1{
     \vskip15true pt plus 4pt minus4pt
     \bgroup
     \bxf
     \noindent
     \raggedright
     \pretolerance=10000
     \ignorespaces #1
     \vskip6true pt plus4pt minus4pt\egroup
     \noindent\ignorespaces}%
 \def \subsubs#1{
     \vskip15true pt plus 4pt minus 4pt
     \bgroup
     \bf
     \noindent
     \ignorespaces #1\unskip.\ \egroup
     \ignorespaces}
\def\footnoterule{\kern-3pt\hrule width 2true cm\kern2.6pt}
\newcount\footcount \footcount=0
\def\advftncnt{\advance\footcount by1\global\footcount=\footcount}
\def\fonote#1{\advftncnt$^{\the\footcount}$\begingroup\petit
       \def\textindent##1{\hang\noindent\hbox
       to\parindent{##1\hss}\ignorespaces}%
\vfootnote{$^{\the\footcount}$}{#1}\endgroup}

\newcount\sterne
\outer\def\byebye{\bigskip\typeset
\sterne=1\ifx\speciali\undefined\else
\bigskip Special caracters created by the author
\loop\smallskip\noindent special character No\number\sterne:
\csname special\romannumeral\sterne\endcsname
\advance\sterne by 1\global\sterne=\sterne
\ifnum\sterne<11\repeat\fi
\vfill\supereject\end}
\def\typeset{\centerline{\petit This article was processed by the author
using the \TeX\ Macropackage from Springer-Verlag.}}

\voffset=0pt

\def\ref#1  {\noindent \hangindent=35.0pt \hangafter=1 {#1} \par}

\chap{Deep radio observation of the gravitational lens candidate
QSO2345+007} 

\noindent
{\bf Alok Ranjan Patnaik}\hfill\break\noindent
Max-Planck-Institut f\"ur Radioastronomie, Auf dem H\"ugel 69, D--53121
Bonn, Germany.
\medskip\noindent
{\bf Peter Schneider}\hfill\break\noindent
Max-Planck-Institut f\"ur Astrophysik, Postfach 1523, D-85740
Garching, Germany
\medskip\noindent
{\bf Ramesh Narayan}\hfill\break\noindent
Harvard-Smithonian Center for Astrophysics, 60 Garden Street,
Cambridge, MA 02138, USA.

\sec{Abstract:}
The double QSO2345+007 comprises two optical components
separated by  7.1\ arcseconds and
is the most prominent `dark matter' gravitational lens candidate.
Despite being known for more than a decade, optical spectroscopy and
imaging have been unable to determine whether this double QSO is a
binary QSO or a gravitational lens system. In this note we report a
deep VLA observation of this system, yielding a map with
a noise level of 8.5\ts $\mu$Jy per beam. We have
a $4\sigma$ detection of a radio source within one arcsecond of the
optical position of the brighter A-component of the QSO, but no significant
detection of any radio counterpart of the B component.  Given that the
flux ratio in the optical waveband is $\sim$3--4, the gravitational
lens hypothesis for this system would predict a radio flux of image B
of $\lesssim 10$\ts $\mu$Jy.  The non-detection of the B component is
thus consistent with, but does not prove, the lens
interpretation.

\sec{1 Introduction}
The double QSO 2345+007, the third known gravitational lens candidate,
was discovered by Weedman et al. (1982) in a spectroscopic quasar
survey.  The two quasars are separated by about 7 arcseconds, which
makes this the QSO lens candidate with the largest image separation.
A and B have similar line profiles and have the same redshift of
$z_{\rm s}=2.15$ to within $15 \pm 20$ km/s.  While this argues in
favour of the lensing hypothesis (Steidel \& Sargent 1991),
nevertheless, there are some differences in the two spectra which
cause legitimate doubts (compare Steidel \& Sargent 1991 with Steidel
\& Sargent 1990) on the lensing interpretation. The suggestion (Nieto
et al.\ 1988) that the B-component is itself a close double with
separation 0\arcsecf4 further complicated the picture, but has not
been verified by other ground-based observations (Weir \& Djorgovski
1991) or HST imaging (Falco 1994).

Deep imaging of the field (Tyson et al. 1986; see also McLeod, Rieke
\& Weedman 1994) has revealed no obvious candidate for a lensing galaxy, 
yielding rather strong limits on the mass-to-light ratio of any
potential lens between the images.  If 2345+007 is indeed a lensing
system, the lens must have rather unusual properties, or must be
hidden at the position of one of the images so as to be outshone by
the QSO. A third possibility is that the lensing is mainly due to a
cluster of galaxies of high mass-to-light ratio, with only a
relatively small galaxy between the images. This possiblity has
received further support by the report (Bonnet et al. 1993) of
coherent distortions of background galaxies around the field of
2345+007, which indicate the presence of a massive galaxy cluster in
this direction.  Fischer et al. (1994) obtained an extremely deep
image of the field surrounding 2345+007 and detected a faint
($B_J=25$\ts mag) galaxy close to image B, as well as a statistically
significant enhancement of faint galaxies around the QSO
images. Mellier et al.\ts (1994) also found an excess of faint blue
galaxies in the field, centred on the shear field obtained by Bonnet
et al.\ts (1993). These imaging results support the lensing hypothesis
for this system. From the absorption line spectra of the two QSO
images (Foltz et. al.\ 1984; Steidel \& Sargent 1990, 1991), the
probable redshift of the deflector is estimated to be $z_{\rm d}\sim
1.49$.
 
2345+007 is a very interesting and important object for gravitational
lensing studies for at least two reasons.  First, the large image
separation implies that either the lens is both massive and compact.
For example, the lensing models considered in Fischer et al.\ts
(1994), assuming $z_{\rm d}\sim 1.49$, either attribute a huge
velocity dispersion ($\approx 850$\ts km/s) to the galaxy near image
B, or, if the galaxy overdensity is taken into account as a cluster at
the same redshift, this cluster must be extremely massive and
compact. Alternatively, there can be an additional deflector at a
different (lower) redshift, as indicated by three redshift
measurements of galaxies in the field (Bonnet et al.\ts 1993). Second,
2345+007 was the first double QSO for which correlations of the
absorption spectra of the two objects were used to put constraints on
the sizes of absorbing clouds (e.g., Foltz et al.\ts 1984; Steidel \&
Sargent 1991).  The results of this analysis depend critically on
whether the object is lensed.

For the above reasons, it is of great interest to confirm the lensing
nature of 2345+007. Whereas the support for the lensing hypothesis
summarized above may appear convincing, `very good gravitational lens
candidates', based on spectroscopic evidence, had in the past to be
reclassified as binary QSOs (e.g., Djorgovski et al.\ts 1987). Also,
there are some spectral differences between the A and B components, e.g.,
the line-to-continuum ratios for various emission lines differ
significantly (Steidel \& Sargent 1991). While such differences
can be attributed to intrinsic variability of the QSO coupled with the
difference in light-travel-time (estimated to be of order a few
years), to gravitational microlensing of the continuum (e.g.,
Wambsganss 1994, and references therein) and/or the broad line region
(Nemiroff 1988; Schneider \& Wambsganss 1990), or extinction in the
lensing galaxy as indicated by color differences between the images,
the possibility that 2345+007 is a true binary QSO cannot at present
be ruled out.

In an attempt to clarify the nature of 2345+007, we have obtained a
deep radio image of this system with the VLA. A detection of
radio counterparts for both components with about the same flux ratio
as in the optical 
would provide a strong argument in favour of the lensing nature of the
system. In addition, we were also motivated by the quest for the
magnification ratio of the two images, assuming the system to be a
lens.  In the optical regime, the flux ratio varies with time and is
different for the optical continuum and the broad emission lines
(e.g., Steidel
\& Sargent 1991). These flux-ratio changes can be explained by
intrinsic variations of the QSO or by microlensing. The radio flux
ratio should be less affected by these effects. 

In Sect.\ts 2 we describe the observations, and in Sect.\ts 3 we
present the results, which are discussed in Sect.\ts 4.

\sec{2 Observations and data reduction}

We observed 2345+007 on 1994 September 25 for 9~hours at 8.4\ts GHz
using the VLA in its B/C configuration with a total bandwidth of
100\ts MHz. The array configuration and the observing frequency
were chosen to optimise sensitivity and resolution ($\sim$2.5\ts arcsec) for the
image separation of 7.1\ts arcsec. The phase centre
of the observations was at RA(J2000): 23 48 19.33 and Dec(J2000): +00 57
18.90 which is located between the A and B images.  3C48 was observed
twice for flux density calibration (assumed flux density of 3.30\ts
Jy). Since self-calibration is not possible for 2345+007, frequent
phase calibration is essential. J0006-0623 was observed for 2 minutes
every 15 minutes to calibrate the instrumental phases.

The data were analysed using the NRAO {\it AIPS} software package. 
The flux density of J0006-0623 was found to be $2.812\pm0.009$Jy.
We used both uniform and natural weighting of the data to make maps
using the program {\it MX}.

\sec{3 Results}

The map made using natural weighting of the data is presented in
Fig.1. The noise in this map is $8.5~\mu$Jy which is close to the
expected thermal noise. The brightest source in the field is a
$80\pm10\mu$Jy  source at RA(J2000): 23 48 15.943 and Dec(J2000): +00
56 11.70 and lies outside the region shown in Fig.\ts 1.
The errors on positions are about 0.3~arcsec in each
coordinate.
The crosses on Fig.1 mark the location of images A and B (derived from
precessing the optical positions as given by Weedman et~al. 1982). We
detect a source with flux density of $35.0\pm10 \mu$Jy at RA(J2000):
23 48 19.597 and Dec(2000): +00 57 21.70 which is within 1 arcsec of
2345+007A (RA 23 48 19.53, Dec +00 57 20.8). We do not detect any
significant emission at the position of 2345+007B (RA 23 48 19.12 Dec +00 57
16.9). 
\xfigure{1}{8.4GHz VLA map of 2345+007 with a resolution of 2.5 arcsec. 
The contour levels plotted in multiples of 16$\mu$Jy/beam are $-2$,
$-1$, 1, 
1.5, 2, 2.5, 3. Negative contours are dashed.  The rms noise in the map
is 8.5$\mu$Jy/beam. The restoring beam is shown on the lower left-hand
corner. The crosses mark the optical positions of the two images.}
{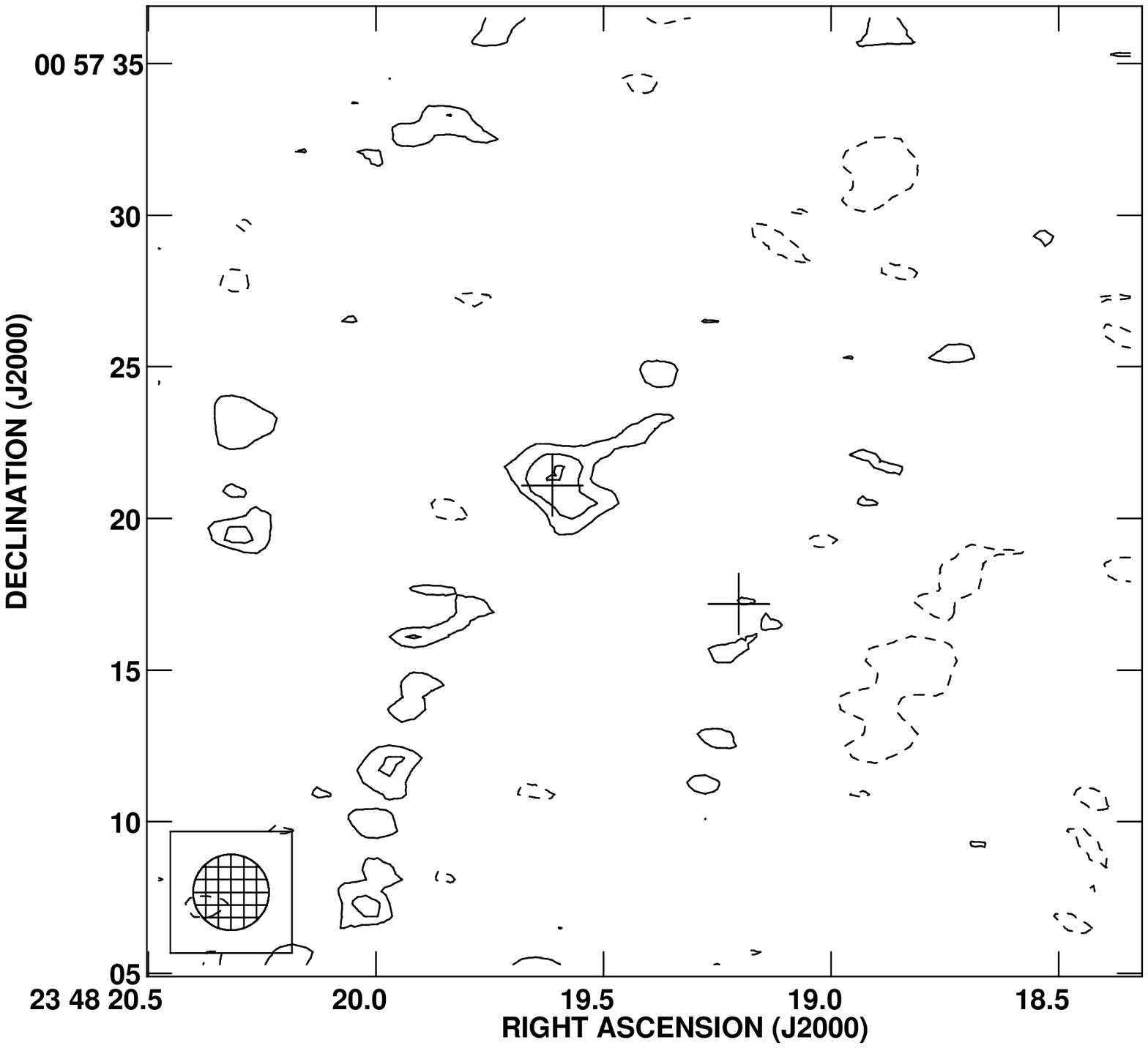}{15}

\sec{4 Discussion}
If 2345+007 is a gravitationally lensed QSO, then for a radio flux of
$S_A=38\pm 10$\ts$\mu$Jy for the A component, we expect the flux of the
B component to be $S_A/r$, where $r=\mu_A/\mu_B$ is the magnification
ratio between the two images.  The magnification ratio can be
estimated from the optical flux ratios, which, however, change in time
and are different for the optical continuum and the emission
lines. The values for $r$ quoted in the literature (see Weir \&
Djorgovski 1991) range from $\sim 2.5$ to $\sim 4.0$. Assuming a value
closer to the upper end of the quoted values, as is justified by the
fact that the MgII-line (which is a low-ionization line supposedly
emitted at fairly large radii within the broad line region and so
least affected by microlensing) indicates $r\sim 4$ (Steidel \&
Sargent 1990), the expected radio flux of image B is about
10\ts$\mu$Jy.  Unfortunately, this is at the noise level of our
maps.  The fact that we do not detect the B component is thus
consistent with the lensing hypothesis.  However, it does not provide
the kind of strong proof we were seeking.

In terms of the radio-to-optical flux ratio, 2345+007A appears to be a
normal radio-quiet QSO. It is well known that about 10\% of
optically-selected QSOs are radio loud, while the rest have fairly
weak radio emission. Kellermann et al.\ts (1989) observed all QSOs
from the Bright Quasar Survey (BQS; Schmidt \& Green 1983) with the
VLA and detected 82\% of the QSOs above a flux level of 200\ts$\mu$Jy.
The BQS QSOs are all brighter than $B\lesssim 16.2$. The A image of
2345+007 is about 3.5\ts mag fainter than the typical QSO from the
BQS.  Therefore, assuming that the radio-to-optical flux ratio does
not strongly depend on the optical brightness of a QSO, we expect the
radio flux density of the A image to be $\sim 50 \mu$Jy.  The
observed radio flux of $35\mu$Jy makes this object quite typical for
its class.  The B image too is then consistent with being a typical
radio-quiet QSO.  

Despite the observational evidence described in the introduction that
2345+007 is a gravitational lens system, some doubts remain.  It is
perhaps better to take a conservative position on this issue since the
implications if the object is truly lensed are substantial.  Even
though a candidate lens galaxy has been observed (Fischer et al.\
1994), its properties and/or the properties of the associated cluster
have to be quite unusual to cause the large observed image
splitting. In particular, if the lens redshift is $z_{\rm d}=1.49$ as
indicated by the rich absorption system at this redshift in the
spectra of the A and B images, it would indicate an enormously massive
and compact mass concentration at quite an early epoch, which would
have implications for current cosmological models. For example, it
is well known that a low-density universe allows the formation of
clusters at higher redshifts than a flat universe (e.g.,
Richstone, Loeb \& Turner 1992; Bartelmann, Ehlers \& Schneider
1993). Also, one of the strongest constraints in mixed cold and hot
dark matter models comes from the presence of compact structures at
early epochs (e.g., Cen \& Ostriker 1994).

Although improving the existing optical and
radio observations of 2345+007 will require a substantial effort, in
view of the important astrophysical implications of the lensing nature
of this object, deeper observations of this spectacular (in terms of
image separation) gravitational lens candidate are worth pursuing.
 
We thank R.\ts Porcas and H.-Th.\ts Janka for carefully reading the
manuscript. \SFB

\sec{References}

\ref{Bartelmann, M., Ehlers, J. \& Schneider, P.\ 1993, A\&A 280, 351.}
\ref{Bonnet, H., Fort, B., Kneib, J.P., Soucail, G. \& Mellier,
Y.\ 1993, A\&A 280, L7.}
\ref{Cen, R. \& Ostriker, J.P.\ 1994, ApJ 431, 451.}
\ref{Djorgovski, S., Perley, R., Meylan, G. \& McCarthy, P.\ 1987, 
ApJ 321, L17.}
\ref{Falco, E.E.\ 1994, in: {\it Gravitational lenses in the
universe}, J.\ Surdej et al. (Eds.), Li\`ege 1994, p.\ts 127.}
\ref{Fischer, P., Tyson, J.A., Bernstein, G.M. \& Guhathakurta, P.\ 1994,
ApJ 431, L71.}
\ref{Foltz, C.B., Weymann, R.J., R\"oser, H.-J. \& Chaffee, F.H. 1984,
ApJ 281, L1.} 
\ref{Kellermann, K.I., Sramek, R., Schmidt, M., Shaffer, D.B. \&
Green, R.\ 1989, AJ 98, 1195.}
\ref{McLeod, B., Rieke, M. \& Weedman, D.\ 1994, ApJ 433, 528.}
\ref{Mellier, Y., Dantel-Fort, M., Fort, B. \& Bonnet, H.\ 1994,
A\&A 289, L15.}
\ref{Nemiroff, R.J.\ 1988, ApJ 335, 593.}
\ref{Nieto, J.-L., et al.\ 1988, ApJ 325, 644.}
\ref{Richstone, D., Loeb, A. \& Turner, E.L.\ 1992, ApJ 393, 477.}
\ref{Schmidt, M. \& Green, R.\ 1983, ApJ 269, 352.}
\ref{Schneider, P. \& Wambsganss, J.\ 1990, A\&A 237, 42.}
\ref{Smette, A. et al.\ 1992, ApJ 389, 39.}
\ref{Steidel, C.C. \& Sargent, W.L.W.\ 1990, AJ 99, 1693.}
\ref{Steidel, C.C. \& Sargent, W.L.W.\ 1991, AJ 102, 1610.}
\ref{Tyson, J.A., Seitzer, P., Weymann, R.J. \& Foltz, C.\
1986, AJ 91, 1274.}
\ref{Wambsganss, J.\ 1994, in: {\it Gravitational lenses in the
universe}, J.\ Surdej et al. (Eds.), Li\`ege 1994, p.\ts 369.}
\ref{Weedman, D.W., Weymann, R.J., Green, R.F. \& Heckman, T.M.\ 1982,
ApJ 255, L5.} 
\ref{Weir, N. \& Djorgovski, S.\ 1991, AJ 101, 66.}

\vfill\eject\end